\pdfoutput=1
\documentclass[usenatbib]{mn2e}
\usepackage{apjfonts}
\usepackage{amssymb}
\usepackage{amsmath}
\usepackage{ctable}
\usepackage{fixltx2e} 

\newcommand{\be}{\begin{equation}}
\newcommand{\ee}{\end{equation}}

\newcommand{\msun}{M_{\sun}}

\newcommand{\paperone}{Paper {\small I}}
\newcommand{\papertwo}{Paper {\small II}}

\newcommand\plotone[2]
 {\centering \leavevmode \includegraphics[width={#2\columnwidth}]{#1}}

\newcommand{\acknowledgments}{\begin{small}\section*{Acknowledgments}\end{small}}
\newcommand\altaffilmark[1]{$^{#1}$}
\newcommand\altaffiltext[1]{$^{#1}$}
\title[SF Criteria in Galaxy Simulations]{The Meaning and Consequences of Star Formation Criteria in Galaxy Models with Resolved Stellar Feedback\vspace{-0.5cm}}

\author[Hopkins et al.]{
\parbox[t]{\textwidth}{ 
Philip F.~Hopkins\thanks{E-mail:phopkins@caltech.edu}\altaffilmark{1,2},
Desika Narayanan\altaffilmark{3}, \&
Norman Murray\altaffilmark{4,5} 
} 
\vspace*{6pt} \\
\altaffiltext{1}{TAPIR, Mailcode 350-17, California Institute of Technology, Pasadena, CA 91125, USA} \\
\altaffiltext{2}{Department of Astronomy and Theoretical Astrophysics Center, University of California Berkeley, Berkeley, CA 94720} \\
\altaffiltext{3}{Steward Observatory, University of Arizona, 933 
N Cherry Ave, Tucson, Az, 85721} \\
\altaffiltext{4}{Canadian Institute for Theoretical Astrophysics, 
60 St.\ George Street, University of Toronto, ON M5S 3H8, Canada} \\
\altaffiltext{5}{Canada Research Chair in Astrophysics} 
\vspace{-0.6cm}
}

\date{Submitted to MNRAS, December, 2012\vspace{-0.6cm}}
\begin{document}
\maketitle
\label{firstpage}

\begin{abstract}
We consider the effects of different criteria for determining where stars will form in gas on galactic scales, in simulations with high (1\,pc) resolution, with explicitly resolved physics of GMC formation and destruction and stellar feedback from supernovae, radiation pressure, stellar winds, and photo-heating. We compare: (1) a self-gravity criterion (based on the local virial parameter and the assumption that self-gravitating gas collapses to high density in a single free-fall time), (2) a fixed density threshold, (3) a molecular-gas law, (4) a temperature threshold, (5) a requirement that the gas be Jeans-unstable, (6) a criteria that cooling times be shorter than dynamical times, and (7) a convergent-flow criterion. We consider all of these in both a MW-like and high-density (starburst or high-redshift) galaxy. With feedback present, all models produce {\em identical} integrated star formation rates (SFRs), in good agreement with the Kennicutt relation; without feedback all produce orders-of-magnitude excessive SFRs. This is totally dependent on feedback and independent of the SF law, even if the ``local'' collapse efficiency is $100\%$. However, the predicted spatial and density distribution depend strongly on the SF criteria. Because cooling rates are generally fast within galaxy disks, and gas is turbulent, criteria (4)-(7) are very ``weak'' and spread the SF uniformly over most of the disk (down to densities $n\sim0.01-0.1\,{\rm cm^{-3}}$). A molecular criterion (3) localizes to slightly higher densities, but still a wide range; for metallicity near solar, it is almost identical to a fixed density threshold at $n\sim1\,{\rm cm^{-3}}$ (well below the mean density in the central MW or starburst systems). A fixed density threshold (2) can always select the highest resolved densities, but must be adjusted both for simulation resolution and individual galaxy properties -- the same threshold that works well in a MW-like simulation will select nearly all gas in a starburst. Binding criteria (1) tend to adaptively select the largest local over-densities, independent of galaxy model or resolution, and automatically predict clustered star formation. We argue that this SF model (possible with other secondary criteria) is most physically-motivated and presents significant numerical advantages in simulations with a large dynamic range.
\end{abstract}

\begin{keywords}
galaxies: formation --- galaxies: evolution --- galaxies: active --- 
star formation: general --- cosmology: theory
\vspace{-1.0cm}
\end{keywords}

\vspace{-1.1cm}
\section{Introduction}
\label{sec:intro}

Modeling star formation accurately is critical for any simulation of galaxy formation. However, cosmological and galaxy-scale simulations still cannot hope to resolve the spatial and density scales on which star formation actually occurs. As a result, simple ``recipes'' must be applied. For example, models typically impose some ``local Schmidt law,'' where gas forms stars at a rate that scales as some power of the density; if the SFR per local free-fall time $t_{\rm ff}\propto 1/\sqrt{G\rho}$ were constant, this would be $\dot{\rho}_{\ast}\propto \rho/t_{\rm ff} \propto \rho^{1.5}$, although in principle any other parameters can be used. 

Usually, applying these models alone would artificially spread star formation among all the gas in the simulation, even cosmologically pristine material at high temperatures. So some additional criteria or restrictions must be included. Most commonly, this amounts to a simple density threshold: $n\gtrsim 0.1\,{\rm cm^{-3}}$ in many cosmological simulations. This is not to say such low-density material directly forms stars: rather, this corresponds crudely to densities where the thermal instability sets in, so some un-resolved fraction of the material (which goes into the sub-grid scaling above) will be able to form stars. Other common requirements include restricting star formation to gas which is below some temperature, or Jeans unstable, or in convergent flows, or which has a short cooling time. Recently, various studies have considered molecular criteria: using some combination of density and metallicity to estimate a sub-grid molecular gas fraction and restricting star formation to the ``molecular'' gas \citep{robertson:2008.molecular.sflaw,
kuhlen:2011.mol.reg.h2.dwarfs}.

These criteria are not trivial. Star formation is observed to be highly clustered under essentially all conditions \citep[][and references therein]{lada:2003.embedded.cluster.review}, and without some criteria such as those above, this is not captured in simulations. This does not just mean that stars are forming in the wrong places. ``Smearing out'' star formation over the disk dramatically suppresses the effects of stellar feedback \citep{governato:2010.dwarf.gal.form}. Massive star clusters allow for e.g.\ overlapping SNe ``bubbles'' that can expand much more efficiently than individual SNe remnants. They also concentrate feedback ``where it is needed,'' i.e.\ it preferentially acts in the dense, star-forming gas. Strong radiation pressure effects arise when photons are trapped in optical thick regions around embedded clusters \citep{hopkins:rad.pressure.sf.fb}. And spreading out star formation leads to spurious geometric cancellation between feedback sources. Without a sufficiently strict minimum SF criterion, the ability of the gas to form realistic phase structure \citep{saitoh:2008.highres.disks.high.sf.thold}, or blow winds that regulate its baryon content \citep{governato:2010.dwarf.gal.form} and form realistic disks \citep{pontzen:2011.cusp.flattening.by.sne,governato04:resolution.fx} can be fundamentally altered. 

Unfortunately, in practice the physical interpretation of these criteria often depends both on the resolved dynamic range of the simulation and on the mean properties of the galaxies being simulated, and to obtain similar results they must be numerically adjusted accordingly. 

However, it is increasingly clear that the ISM is governed by super-sonic turbulence over a wide range of scales. Consider, then, a locally self-gravitating region of the ISM ``supported'' by turbulence. In the absence of some feedback disrupting it or ``pumping'' the dispersion, the turbulent support will be damped in a single crossing time; as a result the region will collapse to arbitrarily high densities in about one free-fall time. This is true even if the energy of contraction maintains a constant virial equilibrium at each radius \citep[][]{hopkins:frag.theory}. At sufficiently high densities, {\em eventually} all of the above criteria must be met; so long as the SFR increases with density, eventually an order-unity fraction of the gas will be consumed into stars. So -- in the absence of some self-regulation -- the time-averaged SFR should be 
$\dot{\rho}_{\ast}\approx \rho/t_{\rm collapse} \sim \rho/t_{\rm ff}$ 
{\em regardless} of the ``true'' local star formation criteria/law. This is precisely what is seen in detailed simulations of turbulent media, in the absence of ``pumping'' to unbind collapsing regions 
\citep{ballesteros-paredes:2011.dens.pdf.vs.selfgrav,
padoan:2011.new.turb.collapse.sims,padoan:2012.sfr.local.virparam}. 

We stress that this does not mean the total SFR, even within a dense parcel of 
gas, will actually be as large as $\rho_{\rm gas}/t_{\rm ff}$. If feedback {\em is} present, 
it can self-regulate. As soon as some gas turns into stars, 
feedback can act and disrupt the bound material, terminating 
the star formation locally and suppressing nearby star formation even in the 
dense gas. In such a model, however, the ``net'' efficiency is actually predicted self-consistently from the feedback model, rather than imposed by the sub-grid model (by, say, forcing some by-hand low normalization of $\dot{\rho_{\ast}}(\rho)$.

Of course, if a simulation resolves a bound region, then this collapse will be followed self-consistently. 
What we require is a mechanism to 
treat further collapse, where it would occur, below our resolution limit. In this paper, we propose a simple adaptive self-gravity criterion for star formation in galaxy-scale and cosmological simulations, motivated by the numerical simulations above. We compare 
it to other common criteria in the literature, and examine the implications for the equilibrium SFRs and both the predicted spatial and density distributions of SF in different galaxy environments.

\vspace{-0.5cm}
\section{A Simple Self-Gravity Criterion}
\label{sec:model}

On some scale $\delta r$, self-gravity requires 
$\sigma_{\rm eff}^{2} + c_{s}^{2} < \beta \,G\,M(<\delta r)/\delta r$, 
where $\sigma_{\rm eff}$ includes the contributions from both 
rotational and random motions and $\beta$ is an appropriate 
constant that depends on the internal structure in $\delta r$.
In practice the gas of interest is always highly super-sonic 
in our simulations (and in observations), so we can ignore the $c_{s}$ term here. 
We then obtain the usual virial parameter 
\be
\alpha \equiv \sigma_{\rm eff}^{2}\,\delta r/\beta\,G\,M(<\delta r). 
\ee
To determine {\em local} binding, we wish to describe 
$\alpha$ in the limit where $\delta r$ is small. 
Then $M(<\delta r) = (4\pi/3)\,\bar{\rho}\,\delta r^{3}$ 
where $\bar{\rho}$ is the average density in $\delta r$, 
and 
$\sigma_{\rm eff} \rightarrow \frac{\delta {\bf v}}{\delta {\bf r}}\,\delta r$, 
or more formally 
\be
\sigma_{\rm eff}^{2} = \beta_{v}\,{\bigl(} |\nabla\cdot {\bf v}|^{2} + |\nabla\times {\bf v}|^{2} {\bigr)}\,\delta r^{2}
\equiv {\Bigl(}\frac{\delta\,v}{\delta\,r}{\Bigr)}^{2}\,\delta r^{2}
\ee
Here the $\nabla\cdot {\bf v}$ term accounts for the local radial velocity dispersion 
and inflow/outflow motions, while the $\nabla\times {\bf v}$ term accounts for 
internal rotational/shear and tangential dispersion.
The $\beta_{v}$ term depends on the internal structure again but is close to 
unity.

Combining these terms, we can derive the 
formally resolution-independent criterion: 
\be 
\label{eqn:alpha}
\alpha \equiv \frac{\beta^{\prime}}{2}\,\frac{|\nabla\cdot {\bf v}|^{2} + |\nabla\times {\bf v}|^{2}}{G\,\rho} < 1
\ee
where $\beta^{\prime}\approx1/2$ collects the order-unity terms above. 
This pre-factor depends on the internal mass profile and velocity structure, but only weakly: 
for e..g\ a Plummer sphere or \citet{hernquist:profile} mass distribution with 
pure isotropic, rotationally supported, or constant velocity gradient orbits the 
range is $\beta^{\prime}\approx0.5-0.6$.

This criterion is well-behaved in the local 
limit and does not explicitly depend on any numerical parameters of the 
simulation (spatial or mass resolution), nor does it require inserting any 
``ad hoc'' threshold or normalization criterion. Because it depends only on 
the local velocity gradient and density, it is trivial to implement in either 
Lagrangian (SPH) or Eulerian (grid) codes (as compared to an explicit 
evaluation of the binding criterion over some {\em resolved} scale length, 
which requires a neighbor search, and can be prohibitively expensive in 
certain situations). 

{\em Implicitly}, the velocity gradients and average density are always evaluated 
at the scale of the resolution limit. If going to higher resolution would change these 
quantities, then of course the criterion would give a different result (but as 
a consequence of the physical, not numerical, difference).

\vspace{-0.7cm}
\section{The Simulations}
\label{sec:sims}

The simulations used here are described in detail in 
\citet{hopkins:rad.pressure.sf.fb} 
(hereafter \paperone; see \S~2 \&\ Tables~1-3) and 
\citet{hopkins:fb.ism.prop} (\papertwo; \S~2).
We briefly summarize the most important properties here. 
The simulations were performed with the parallel TreeSPH code {\small 
GADGET-3} \citep{springel:gadget}. They include stars, dark matter, and gas, 
with cooling, shocks, star formation, and stellar feedback. 

\vspace{-0.5cm}
\subsection{Star Formation Criteria}
\label{sec:sims:cooling.sf}

Star formation is allowed only in gas that meets some set of criteria, 
for example in density or temperature. 
Within the gas that is flagged as ``star forming,'' our standard model assumes 
$\dot{\rho}_{\ast} = \epsilon\,{\rho}/{t_{\rm ff}}$
where $t_{\rm ff}$ is the free-fall time 
and $\epsilon$ is some efficiency. Unless otherwise specified, we 
set $\epsilon=0.015$, to match the average observed efficiency in 
dense gas \citep[e.g.][and references therein]{krumholz:sf.eff.in.clouds}.

There are several criteria that can be imposed to determine the 
gas allowed to form stars: 

{\bf (1) Self-Gravity:} We require a region be {\rm locally} self-gravitating as 
described in \S~\ref{sec:model}, i.e.\ $\alpha<1$. 
But these regions are assumed to collapse in a single free-fall time, 
so $\epsilon=1$ (this approximates the results in individual cloud simulations of \citealt{padoan:2012.sfr.local.virparam}).

{\bf (2) Density:} Star formation is allowed above a simple density threshold $n>n_{0}$, 
where we adopt $n_{0} = 100\,{\rm cm^{-3}}$ to ensure this selects only over-dense 
gas inside of typical GMCs.

{\bf (3) Molecular Gas:} We calculate the molecular fraction 
$f_{\rm H_{2}}$ of all gas as a function of the local column density and 
metallicity following \citet{krumholz:2011.molecular.prescription} and allow star formation 
only from the molecular gas (i.e.\ multiply $\epsilon$ by $f_{\rm H_{2}}$). 

{\bf (4) Temperature:} We allow star formation only below a minimum temperature 
$T<T_{\rm min}$. Here we adopt $T_{\rm min}=100\,K$, (chosen to approach the minimum 
temperatures the simulation can resolve). At these temperatures, we expect this 
to be very similar to criterion {\bf (2)}.

{\bf (5) Jeans Instability:} We require the gas be locally Jeans-unstable below the 
resolution limit: $c_{s} < h_{\rm sml}\,\sqrt{4\pi\,G\,\rho}$ ($h_{\rm sml}$ is the SPH smoothing length). Given the Lagrangian 
nature of the simulations, this translates to a temperature  
$T\lesssim 100\,{\rm K}\,(m_{i}/300\,\msun)/(h_{\rm sml}/{\rm 10 pc})$ where $m_{i}$ is the particle 
mass. 

{\bf (6) Converging Flows:} Star formation is allowed only in convergent flows, 
i.e.\ where $\nabla \cdot {\bf v} < 0$.

{\bf (7) Rapid Cooling:} We allow star formation only in regions 
where the cooling time is less than the dynamical time, 
$t_{\rm cool} < 1/\sqrt{G\,\rho}$. 

\vspace{-0.5cm}
\subsection{Cooling \&\ Feedback}
\label{sec:sims:feedback}

Gas follows an atomic cooling curve with additional fine-structure 
cooling to $\sim10\,$K. At all the scales we resolve, the cooling time 
in dense gas tends to be much shorter than the dynamical time for any 
temperatures $T\gtrsim10^{4}\,$K 
where the thermal pressure would be significant, 
and the minimum resolved scales are significantly larger than the sonic length. 
As a result, varying the cooling curve shape, magnitude, or 
metallicity dependence within an order of magnitude has no significant effect on 
any of our conclusions. 

Stellar feedback is included, from a variety of mechanisms.

(1) {Local Momentum-Driven Winds} from Radiation Pressure, 
Supernovae, \&\ Stellar Winds: Gas within a GMC (identified 
with an on-the-fly friends-of-friends algorithm) receives a direct 
momentum flux from the stars in that cluster/clump. 
The momentum flux is $\dot{P}=\dot{P}_{\rm SNe}+\dot{P}_{\rm w}+\dot{P}_{\rm rad}$, 
where the separate terms represent the direct momentum flux of 
SNe ejecta, stellar winds, and radiation pressure. 
The first two are directly tabulated for a single stellar population as a function of age 
and metallicity $Z$ and the flux is directed away from the stellar center. 
Because this is interior to clouds, the systems are always optically thick, so the 
latter is approximately $\dot{P}_{\rm rad}\approx (1+\tau_{\rm IR})\,L_{\rm incident}/c$, 
where $1+\tau_{\rm IR} = 1+\Sigma_{\rm gas}\,\kappa_{\rm IR}$ accounts 
for the absorption of the initial UV/optical flux and multiple scatterings of the 
IR flux if the region is optically thick in the IR (with $\Sigma_{\rm gas}$ calculated 
for each particle). 

(2) {Supernova Shock-Heating}: Gas shocked by 
supernovae can be heated to high temperatures. 
We tabulate the SNe Type-I and Type-II rates from 
\citet{mannucci:2006.snIa.rates} and STARBURST99, respectively, as a function of age and 
metallicity for all star particles and stochastically determine at 
each timestep if a SNe occurs. If so, the appropriate mechanical luminosity is 
injected as thermal energy in the gas within a smoothing length of the star particle. 

(3) {Gas Recycling and Shock-Heating in Stellar Winds:} Gas mass is returned 
to the ISM from stellar evolution, at a rate tabulated from SNe and stellar mass 
loss (integrated fraction $\approx0.3$). The SNe heating is described above. Similarly, stellar winds 
are assumed to shock locally and inject the appropriate tabulated mechanical 
luminosity $L(t,\,Z)$ as a function of age and metallicity into the gas within a smoothing length. 

(4) {Photo-Heating of HII Regions and Photo-Electric Heating}: We also tabulate the rate of production of ionizing photons for 
each star particle; moving radially outwards from the star, we then ionize each neutral gas particle (using 
its density and state to determine the necessary photon number) 
until the photon budget is exhausted. Ionized gas is maintained at a minimum $\sim10^{4}\,$K until 
it falls outside an HII region. Photo-electric heating is followed in a similar manner using the heating rates from \citet{wolfire:1995.neutral.ism.phases}.

(5) {Long-Range Radiation Pressure:} Photons which escape the local GMC (not 
accounted for in (1)) can be absorbed at larger radii. Knowing the intrinsic SED of each star 
particle, we attenuate integrating the local gas density and gradients to convergence. 
The resulting ``escaped'' SED gives a flux that propagates to large distances, and 
can be treated in the same manner as the gravity tree to give the local net incident flux 
on a gas particle. The local absorption is then calculated integrating over a frequency-dependent 
opacity that scales with metallicity, and the radiation pressure force is imparted. 

Details and numerical tests of these models are discussed in \papertwo. 
All energy, mass, and momentum-injection rates are taken as-is from the stellar 
population models in STARBURST99, assuming a \citet{kroupa:imf} IMF, without any free parameters.
Subtle variations in the implementation do not make significant differences to our conclusions. 
Most important, we do {\em not} ``turn off'' or otherwise alter any of the cooling or hydrodynamics 
of the gas.

\vspace{-0.5cm}
\subsection{Galaxy Models}
\label{sec:sims:galaxies}

We implement the model in two distinct initial disk models, 
chosen to span a wide range in ISM densities.
Each has a bulge, stellar and gaseous disk, halo, and central BH (although to isolate the 
role of stellar feedback, models for BH growth and feedback are disabled). 
At our standard resolution, each model has $\sim 0.3-1\times10^{8}$ total particles, 
giving particle masses of $\sim500\,\msun$ and typical $\sim5$\,pc smoothing lengths in the dense gas\footnote{Smoothing lengths are set adaptively as in \citet{springel:entropy}, with an approximately constant $64$ neighbors enclosed within the smoothing kernel.}, 
and are run for a few orbital times each. 
The disk models include: 

(1) MW: a MW-like galaxy, with 
with baryonic mass $M_{\rm bar}=7.1\times10^{10}\,\msun$ 
(gas $m_{g}=0.9\times10^{10}\,\msun$, bulge $M_{b}=1.5\times10^{10}\,\msun$, 
the remainder in a stellar disk $m_{d}$) and halo mass $M_{\rm halo}=1.6\times10^{12}\,\msun$. 
The gas (stellar) scale length is $h_{g}=6.0\,$kpc ($h_{d}=3.0$). At standard resolution, 
gas particles and new stars have mass $m\approx500\,\msun$ and the force softening 
$\epsilon\approx 4\,$pc.

(2) HiZ/Starburst: a massive starburst disk with densities typical of the central couple 
kpc in low-$z$ galaxy mergers, or the larger-scale ISM in star-forming galaxies at $z\sim2-4$.
Here $M_{\rm halo}=1.4\times10^{12}\,\msun$ 
and baryonic 
$(M_{\rm bar},\,m_{b},\,m_{d},\,m_{g})=(10.7,\,0.7,\,3,\,7)\times10^{10}\,\msun$ 
with scale-lengths 
$(h_{d},\,h_{g})=(1.6,\,3.2)\,{\rm kpc}$.

\begin{figure}
    \centering
    \plotone{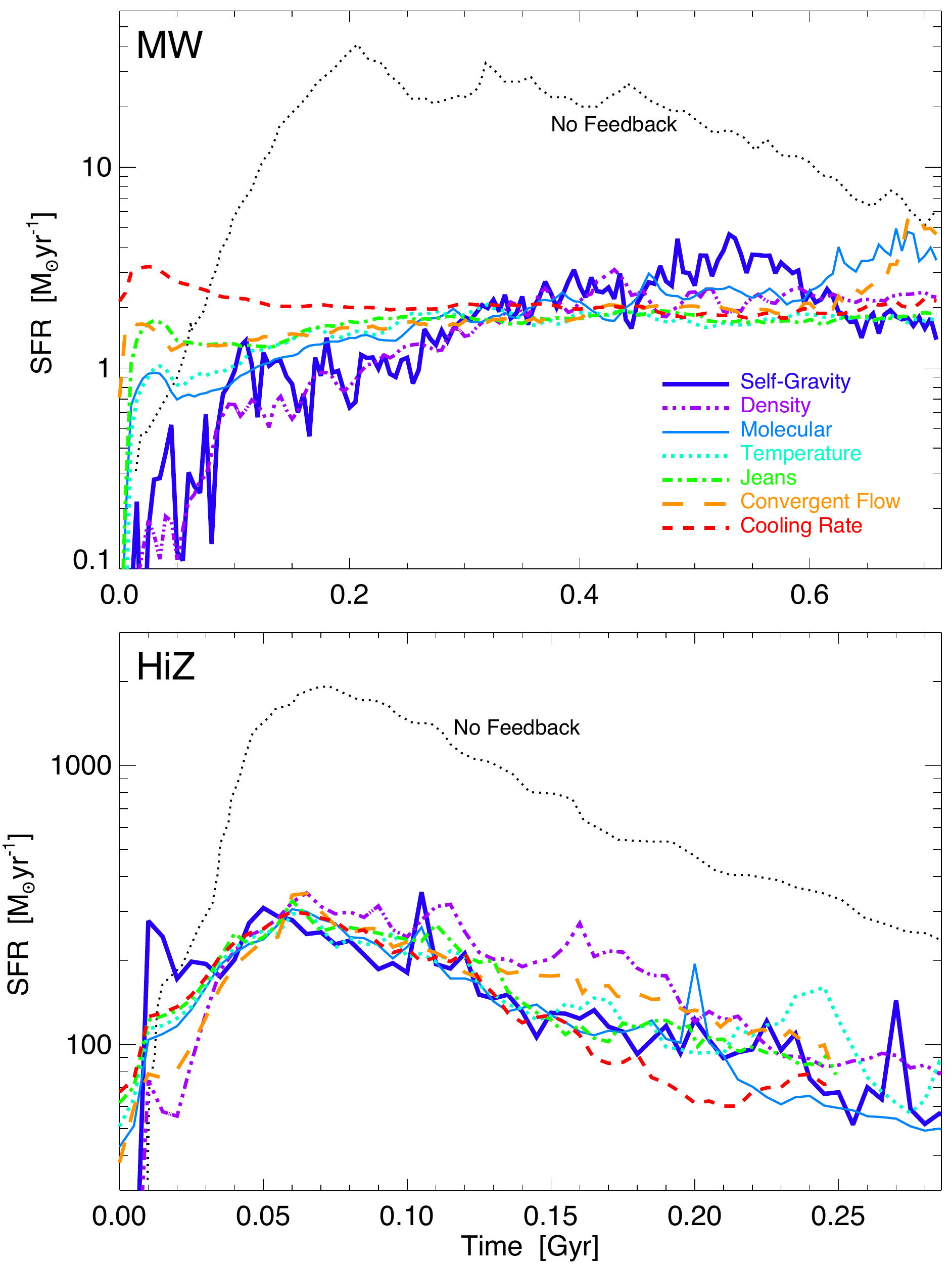}{0.9}
    \caption{SFR vs.\ time for the simulations with different star formation prescriptions (as labeled), 
    in the MW-like ({\em top}) and starburst/HiZ ({\em bottom}) disk models.
    The SF prescription has no effect on the actual SFRs (model 
    differences, after the first couple dynamical times where non-equilibrium effects appear, 
    are consistent with random variation). With realistic stellar feedback 
    models, the SFR is entirely set by feedback (the mass in young stars needed 
    to prevent runaway collapse). With no feedback, SF efficiencies are extremely large (SFRs 
    much larger than observed); with feedback, the SFRs agree well with observations 
    in both models. In particular, despite having an instantaneous, local efficiency $\epsilon=1$ 
    in bound clouds, the actual average efficiency $\langle \epsilon \rangle$ of the ``self-gravity'' 
    model is a realistic $\sim1-5\%$.
    \label{fig:sfh}}
\end{figure}

\begin{figure}
    \centering
    \plotone{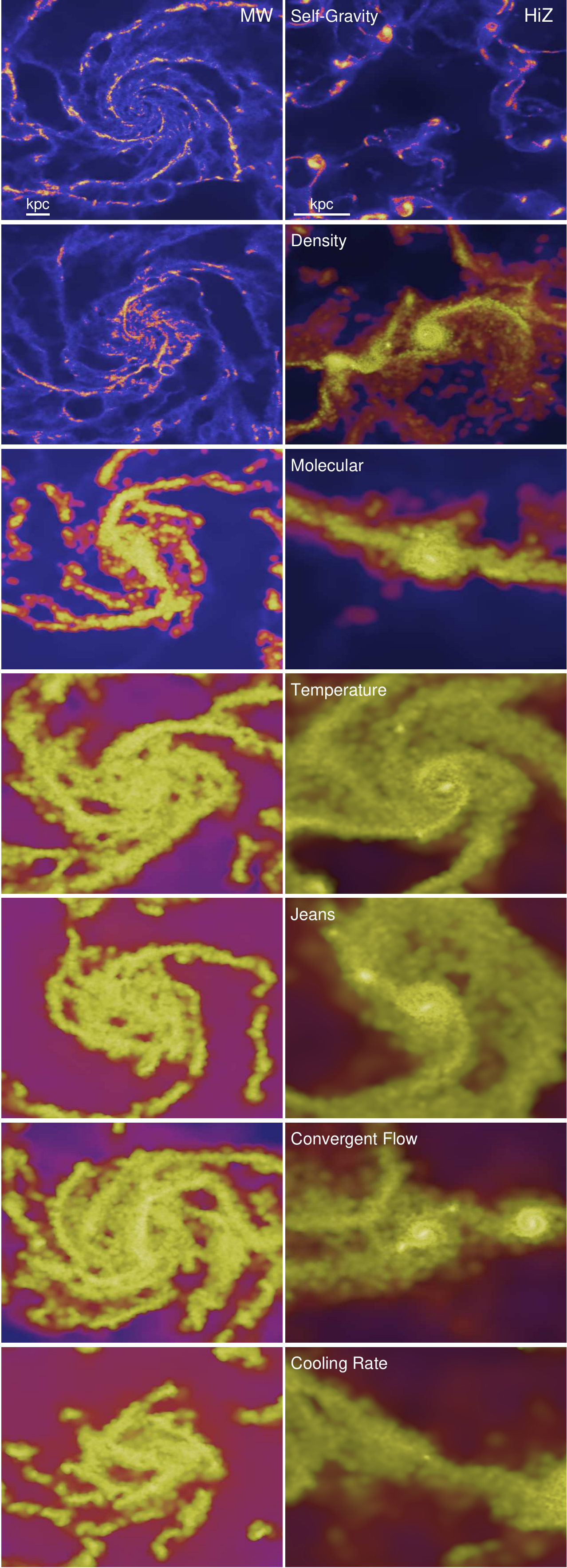}{0.88}
    \caption{Gas surface density (intensity), with star-forming 
    regions color-coded as red-yellow (with increasing specific SFR).
    The {\em distribution} of SF varies with SF law.
    Local binding criteria select locally over-dense regions in all cases.
    A fixed density threshold works well for most of the MW disk, but fails 
    in the central, high-$\rho$ regions of the HiZ model (where the mean 
    density is above-threshold).
    A molecular law works reasonably well 
    in the outer regions of the MW model, but ``smears'' SF among 
    a much wider range of gas in spiral arms; in the HiZ/starburst nucleus model it  
    identifies all gas as molecular.
    Temperature, Jeans stability, cooling rate, 
    and convergent-flow criteria select gas at nearly all densities.    
    \label{fig:distrib}}
\end{figure}

\begin{figure}
    \centering
    \plotone{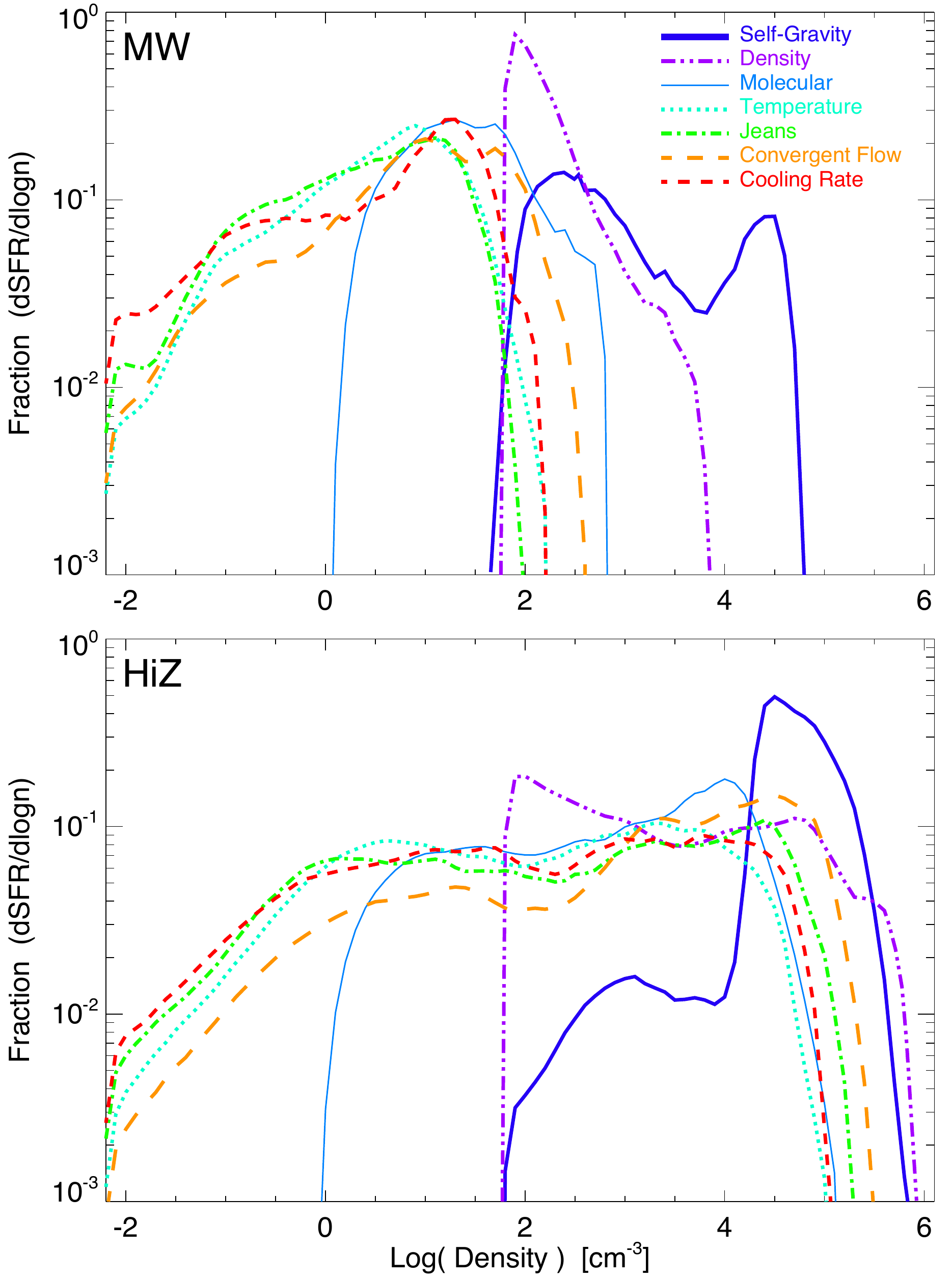}{1.}
    \caption{SFR-weighted density distribution for the simulations with 
    different star formation prescriptions (labeled) in the MW-like ({\em top}) and 
    starburst/HiZ ({\em bottom}) disk models.
    The self-gravity criterion identifies the most dense regions.
    A fixed-density threshold simulation is dominated by SF near the threshold, 
    which is much lower than the maximum densities in the HiZ model. 
    A molecular criterion effectively corresponds to a much lower threshold; for the metallicities 
    here $\approx 1\,{\rm cm^{-3}}$. 
    The other criteria spread the SF across almost all the gas, even at very low densities.
    \label{fig:rho}}
\end{figure}

\vspace{-0.5cm}
\section{Results \&\ Discussion}
\label{sec:discussion}

Figure~\ref{fig:sfh} shows the total SFR versus time for the models, with feedback 
enabled and disabled. We immediately see that the total SFR, hence the {\em net} 
SF efficiency, is almost totally independent of the SF criteria. However, it depends 
dramatically on feedback. 
With feedback on, the global SFR is $\dot{M}_{\ast}\sim0.02\,M_{\rm gas}/t_{\rm dyn}$, 
in agreement with observations. Without feedback, it is $\dot{M}_{\ast}\sim 0.5\,M_{\rm gas}/t_{\rm dyn}$.\footnote{The apparent ``convergence'' of the feedback and no-feedback runs at late times occurs because the no-feedback runs exhaust all their gas in just a few dynamical times. Thus the SFR declines in absolute terms, even though it remains fixed in units of $M_{\rm gas}/t_{\rm dyn}$. This is shown explicitly in \paperone\ \&\ \papertwo. We also refer interested readers to \paperone, Fig.~11 for an explicit comparison of the simulations and the observed Schmidt-Kennicutt relation.}
Regardless of {\em how} mass turns into stars, a certain feedback strength, hence mass 
in young stars, is needed to inject enough momentum to offset dissipation and 
prevent runaway collapse. Because cooling is rapid, the galaxy can always find a way to 
get ``enough'' gas to the relevant densities/temperatures/binding criteria -- 
at which point the SFR is a function of feedback efficiency alone. 
This is discussed in much greater detail in \paperone, where we show 
that, for example, changing the local SF law (efficiencies, power-law indices, or density thresholds) 
makes no difference here, but changing the feedback strength directly changes the net 
efficiency. Without explicit feedback models, a local SF model that turned self-gravitating 
gas into stars with an efficiency $\sim1$ would grossly over-produce observed 
SFRs; however, once feedback is included, the {\em mean} SF efficiency even in dense 
gas is a self-regulating quantity and we can safely consider such a prescription.\footnote{In \papertwo, we examine how different feedback mechanisms individually affect the SF efficiency. In the lower-density regime typical of the MW-like model, the dominant mechanism is a combination of photoionization heating (and resulting warm gas pressure) and the momentum injection in overlapping SNe explosions. In the higher-density regime of the HiZ model, it is predominantly radiation pressure on dusty gas from the nearly Eddington-limited starburst.}

The SF criteria does, however, affect the 
the spatial and density distribution of star formation in each model, 
shown in Figures~\ref{fig:distrib}-\ref{fig:rho}. 
With arbitrarily high resolution, we would ideally want all the star formation to 
be highly spatially clustered, and concentrated in absolute densities of 
$\gtrsim10^{6}\,{\rm cm^{-3}}$. Resolution limits mean this cannot be realized; 
the ``next-best'' aim of these star formation criteria is to identify the highest 
{\em local} over densities and highest resolvable densities, which will 
contain sub-regions that collapse further, and associate star formation with those 
regions. 

Our proposed self-gravity criterion {\bf (1)} adaptively selects the over-dense regions 
in all galaxy models, giving a realistic ``clumpy'' and clustered morphology 
for the star formation. In the HiZ model in particular, it is able to associate 
star formation with a number of small sub-clumps in the central regions, with densities 
$\gtrsim10^{4}\,{\rm cm^{-3}}$. (Note that the apparent ``bimodality'' here is artificial, caused by a bottleneck at the highest densities allowed by our numerical resolution).

A pure density criterion {\bf (2)} works well when the threshold density is much 
larger than the background mean; however, it qualitatively changes character 
when the threshold falls below the mean density (the central regions of the HiZ model).
Then, obviously, the SF becomes ``smeared'' uniformly across all the gas, 
defeating the purpose of a density threshold in the first place. This is 
clear in the morphology in Fig.~\ref{fig:distrib}, where the ``clumps'' 
previously evident have been largely ``smeared out'' by more extended star formation.
Likewise in Fig.~\ref{fig:rho}, the SFR is dominated by gas near the threshold 
density, even though it is clear that densities up to $4\,$dex larger can be resolved.
A number of authors have shown that this leads to unphysical 
distributions of stars and star formation, and can artificially suppress 
the efficiency of feedback as well (see \S~\ref{sec:intro}). 
We stress that at least some of the cases here cannot simply be resolved 
by raising the threshold further: for example, in simulating galaxy nuclei, the 
minimum density any clump must have simply to avoid tidal disruption (a clear 
requirement for star formation) scales $\propto r^{-3}$, 
where $r$ is the distance to the BH. Thus properly selecting ``overdensities'' 
at $\sim1\,$pc would require a threshold a factor of $\sim10^{9}$ larger 
than the threshold at $\sim$kpc -- and the threshold we use on those scales 
is already much larger than the value chosen in 
most simulations!

Identifying star formation with molecules {\bf (3)} is reasonably similar 
to the choice of a high threshold density or self-gravity, when we focus on 
regions of very low mean surface density/opacity. 
When the average surface density is $\lesssim 10\,\msun\,{\rm pc^{-2}}$ (at 
$Z\sim Z_{\sun}$), 
the medium is not self-shielding and becomes atomic-dominated, so these criteria all similarly select 
overdense regions where rapid cooling has enabled collapse. 
But as soon as the density rises much above this value, 
the (dense) gas is essentially {\em all} molecular, and the criterion becomes 
meaningless (distributing star formation equally among all gas). 
This ``smears out'' star formation in the dense gas, evident in Figure~\ref{fig:distrib} (which 
now appears as if it were effectively lower-resolution). 
In fact, for metallicities of $\gtrsim 0.1\,Z_{\sun}$, 
we see in Figure~\ref{fig:rho} that this criterion is nearly identical to invoking a 
relatively low ``threshold'' density of $n_{0}\sim 1\,{\rm cm^{-3}}$.
At lower metallicities significant differences appear 
\citep{kuhlen:2011.mol.reg.h2.dwarfs}, but there is almost no difference in an 
{\em instantaneous} sense between this model and a threshold density -- the 
differences owe to the fact that at these metallicities cooling rates are sufficiently 
suppressed such that the cooling time is no longer short compared to the 
dynamical time \citep{glover:2011.molecules.not.needed.for.sf}. 

A temperature criterion {\bf (4)},  
Jeans criterion {\bf (5)}, 
and rapid cooling criterion {\bf (7)} 
identify the ``molecular'' gas 
from criterion {\bf (3)}, but also include a wide range of 
gas at even lower densities. 
This can include e.g.\ adiabatically 
cooled gas in winds and cold clumps which have been shredded 
by feedback. As a result SF in Figure~\ref{fig:distrib} is effectively distributed over 
all the gas (clearly over several dex in Fig.~\ref{fig:rho}, down to $n\sim0.01-0.1\,{\rm cm^{-3}}$); 
the only reason it is concentrated towards the center at all is because 
of the global concentration of gas mass.

An inflow/convergent flow criterion {\bf (6)} has almost no 
effect on the distribution of star formation, except to 
randomly select $\sim1/2$ of the gas. 
This is because the medium is turbulent on all scales, 
so ${\rm SIGN}(\nabla\cdot{\bf v})$ is basically random.


The binding energy criterion used here appears to have a number of 
advantages over the traditional fixed density threshold criterion for following 
star formation in galaxy simulations. It is physically well-motivated and 
removes the ambiguity associated with assigning a specific density -- it also agrees well with the results of much higher-resolution simulations of star formation in individual local regions \citep{padoan:2012.sfr.local.virparam}. 
With realistic feedback models present, it is possible to set $\epsilon=1$, i.e.\ have 
order-unity efficiency in bound regions, and correctly reproduce the observed 
SF efficiencies both globally and in dense gas. This removes the dependence of 
the observed Schmidt-Kennicutt relation on all resolved scales on the ``by hand'' 
insertion of a specific efficiency. Interestingly, the total mass in dense ($n\gtrsim100\,{\rm cm^{-3}}$) gas is similar in the runs with a density threshold and $\epsilon=0.015$, and those with a self-gravity criterion and $\epsilon=1$. In the latter case, a large fraction of the dense gas at any moment is not locally self-gravitating; the broad distribution of virial parameters in simulated GMCs is shown in \papertwo, Figs.~17-18 \&\ \citealt{dobbs:2011.why.gmcs.unbound}. So in this (local) sense as well as in the global average SFR shown here, the ``average efficiencies'' emerge similarly. A more quantitative calculation of the dense gas distributions as tracers of the local star formation efficiency is presented in \citet{hopkins:dense.gas.tracers}. 

Most importantly, the self-gravity criterion is inherently adaptive, and so allows 
for the simultaneous treatment of a wide dynamic range -- critical for simulations of 
e.g.\ galaxy mergers, active galactic nuclei, and high-resolution cosmological simulations. 
In these models, gas which is likely to be ``star-forming'' in one context (say a GMC in the 
outer parts of a galaxy disk) might be orders-of-magnitude below the densities needed 
for it to be even tidally bound in other regimes, so no single 
density threshold is practical for realistic simulation resolution limitations. For example, 
in the Milky Way, the central molecular zone is observed to have very high gas densities relative to the solar neighborhood and even many GMCs, but is not strongly self-gravitating, and so appears to have a SFR far lower than what would be predicted by a simple density threshold argument \citep[see][]{longmore:2013.mw.cmz.sf.thresholds}.

Of course it is possible to combine the criteria here, requiring self-gravity 
in addition to some density/temperature/molecular threshold, for example. 
However, based on our results, the addition of 
most of these criteria will not dramatically modify the results from a binding 
criterion alone. Moreover, it is not entirely obvious if they add physical information -- 
at low densities, for example, one might posit that a region should not form stars 
unless it is also cold. But if it does not form stars (and so cannot be disrupted by feedback), 
it would collapse (if it could be resolved) to much higher densities, 
at which point it should eventually become cold and molecular as well. 
The exception is the cooling time criterion -- rapid collapse 
implicitly assumes efficient cooling; it is less clear what will happen to a clump 
that is bound but cannot dissipate. Likewise, one may wish to adopt other criteria 
for non-bound regions. For example, associating a low but non-zero efficiency even with un-bound but  turbulent regions above some density threshold, to represent the fact that there is an (unresolved) distribution of densities therein, some of which might be self-gravitating themselves \citep[see e.g.][]{krumholz.schmidt,hopkins:excursion.ism}. We have experimented with such a prescription (with $\epsilon=0.01$ for $n_{\rm crit}>100\,{\rm cm^{-3}}$), and find the contribution from the un-bound material is generally sub-dominant (but non-negligible). 

Ultimately, testing which prescription is most accurate in simulations should involve direct comparison with observations. Since we have shown that the different simulation criteria are degenerate in their predictions of the total SFR and star formation ``efficiency,'' this is not a good observational constraint to test the models. However, it is possible to measure how the observed SF in resolved galaxies is distributed with respect to the average gas density (or column density) on different {\em resolved} scales, essentially constructing a direct analog of Fig.~\ref{fig:rho}. The differences there and in the spatial distribution of SF in Fig.~\ref{fig:distrib} should also be manifest in quantities such as the spatial correlation function of star formation (and young stars), and the sizes of different star-forming regions (e.g.\ the characteristic sizes of the star-forming portions of spiral arms or GMC complexes). The behavior of the local, small-scale star formation law (as a function of density or other parameters) also ultimately informs the ``sub-grid'' physics these models are intended to reproduce.

\vspace{-0.7cm}
\acknowledgments 
Support for PFH was provided by NASA through Einstein Postdoctoral Fellowship Award Number PF1-120083 issued by the Chandra X-ray Observatory Center, which is operated by the Smithsonian Astrophysical Observatory for and on behalf of NASA under contract NAS8-03060. NM is supported in part by NSERC and by the Canada Research Chairs program. \\

\bibliography{/Users/phopkins/Documents/work/papers/ms}

\begin{thebibliography}{27}
\expandafter\ifx\csname natexlab\endcsname\relax\def\natexlab#1{#1}\fi

\bibitem[{{Ballesteros-Paredes} {et~al.}(2011){Ballesteros-Paredes},
  {Vazquez-Semadeni}, {Gazol}, {Hartmann}, {Heitsch}, \&
  {Colin}}]{ballesteros-paredes:2011.dens.pdf.vs.selfgrav}
{Ballesteros-Paredes}, J., {Vazquez-Semadeni}, E., {Gazol}, A., {Hartmann},
  L.~W., {Heitsch}, F., \& {Colin}, P. 2011, \mnras, 416, 1436

\bibitem[{{Dobbs} {et~al.}(2011){Dobbs}, {Burkert}, \&
  {Pringle}}]{dobbs:2011.why.gmcs.unbound}
{Dobbs}, C.~L., {Burkert}, A., \& {Pringle}, J.~E. 2011, \mnras, 413, 528

\bibitem[{{Glover} \& {Clark}(2012)}]{glover:2011.molecules.not.needed.for.sf}
{Glover}, S.~C.~O., \& {Clark}, P.~C. 2012, \mnras, 421, 9

\bibitem[{{Governato} {et~al.}(2004)}]{governato04:resolution.fx}
{Governato}, F., {et~al.} 2004, \apj, 607, 688

\bibitem[{{Governato} {et~al.}(2010)}]{governato:2010.dwarf.gal.form}
---. 2010, \nat, 463, 203

\bibitem[{{Hernquist}(1990)}]{hernquist:profile}
{Hernquist}, L. 1990, \apj, 356, 359

\bibitem[{{Hopkins}(2012)}]{hopkins:excursion.ism}
{Hopkins}, P.~F. 2012, \mnras, 423, 2016

\bibitem[{{Hopkins}(2013)}]{hopkins:frag.theory}
---. 2013, \mnras, 430, 1653

\bibitem[{{Hopkins} {et~al.}(2012{\natexlab{a}}){Hopkins}, {Narayanan},
  {Quataert}, \& {Murray}}]{hopkins:dense.gas.tracers}
{Hopkins}, P.~F., {Narayanan}, D., {Quataert}, E., \& {Murray}, N.
  2012{\natexlab{a}}, \mnras, in press, arXiv:1209.0459

\bibitem[{{Hopkins} {et~al.}(2011){Hopkins}, {Quataert}, \&
  {Murray}}]{hopkins:rad.pressure.sf.fb}
{Hopkins}, P.~F., {Quataert}, E., \& {Murray}, N. 2011, \mnras, 417, 950

\bibitem[{{Hopkins} {et~al.}(2012{\natexlab{b}}){Hopkins}, {Quataert}, \&
  {Murray}}]{hopkins:fb.ism.prop}
---. 2012{\natexlab{b}}, \mnras, 421, 3488

\bibitem[{{Kroupa}(2002)}]{kroupa:imf}
{Kroupa}, P. 2002, Science, 295, 82

\bibitem[{{Krumholz} \& {Gnedin}(2011)}]{krumholz:2011.molecular.prescription}
{Krumholz}, M.~R., \& {Gnedin}, N.~Y. 2011, \apj, 729, 36

\bibitem[{{Krumholz} \& {McKee}(2005)}]{krumholz.schmidt}
{Krumholz}, M.~R., \& {McKee}, C.~F. 2005, \apj, 630, 250

\bibitem[{{Krumholz} \& {Tan}(2007)}]{krumholz:sf.eff.in.clouds}
{Krumholz}, M.~R., \& {Tan}, J.~C. 2007, \apj, 654, 304

\bibitem[{{Kuhlen} {et~al.}(2012){Kuhlen}, {Krumholz}, {Madau}, {Smith}, \&
  {Wise}}]{kuhlen:2011.mol.reg.h2.dwarfs}
{Kuhlen}, M., {Krumholz}, M.~R., {Madau}, P., {Smith}, B.~D., \& {Wise}, J.
  2012, \apj, 749, 36

\bibitem[{{Lada} \& {Lada}(2003)}]{lada:2003.embedded.cluster.review}
{Lada}, C.~J., \& {Lada}, E.~A. 2003, \araa, 41, 57

\bibitem[{{Longmore} {et~al.}(2013)}]{longmore:2013.mw.cmz.sf.thresholds}
{Longmore}, S.~N., {et~al.} 2013, \mnras, 429, 987

\bibitem[{{Mannucci} {et~al.}(2006){Mannucci}, {Della Valle}, \&
  {Panagia}}]{mannucci:2006.snIa.rates}
{Mannucci}, F., {Della Valle}, M., \& {Panagia}, N. 2006, \mnras, 370, 773

\bibitem[{{Padoan} {et~al.}(2012){Padoan}, {Haugb{\o}lle}, \&
  {Nordlund}}]{padoan:2012.sfr.local.virparam}
{Padoan}, P., {Haugb{\o}lle}, T., \& {Nordlund}, {\AA}. 2012, \apjl, 759, L27

\bibitem[{{Padoan} \& {Nordlund}(2011)}]{padoan:2011.new.turb.collapse.sims}
{Padoan}, P., \& {Nordlund}, {\AA}. 2011, \apj, 730, 40

\bibitem[{{Pontzen} \& {Governato}(2012)}]{pontzen:2011.cusp.flattening.by.sne}
{Pontzen}, A., \& {Governato}, F. 2012, \mnras, 421, 3464

\bibitem[{{Robertson} \& {Kravtsov}(2008)}]{robertson:2008.molecular.sflaw}
{Robertson}, B.~E., \& {Kravtsov}, A.~V. 2008, \apj, 680, 1083

\bibitem[{{Saitoh} {et~al.}(2008){Saitoh}, {Daisaka}, {Kokubo}, {Makino},
  {Okamoto}, {Tomisaka}, {Wada}, \&
  {Yoshida}}]{saitoh:2008.highres.disks.high.sf.thold}
{Saitoh}, T.~R., {Daisaka}, H., {Kokubo}, E., {Makino}, J., {Okamoto}, T.,
  {Tomisaka}, K., {Wada}, K., \& {Yoshida}, N. 2008, \pasj, 60, 667

\bibitem[{{Springel}(2005)}]{springel:gadget}
{Springel}, V. 2005, \mnras, 364, 1105

\bibitem[{{Springel} \& {Hernquist}(2002)}]{springel:entropy}
{Springel}, V., \& {Hernquist}, L. 2002, \mnras, 333, 649

\bibitem[{{Wolfire} {et~al.}(1995){Wolfire}, {Hollenbach}, {McKee}, {Tielens},
  \& {Bakes}}]{wolfire:1995.neutral.ism.phases}
{Wolfire}, M.~G., {Hollenbach}, D., {McKee}, C.~F., {Tielens}, A.~G.~G.~M., \&
  {Bakes}, E.~L.~O. 1995, \apj, 443, 152

\end{thebibliography}

\end{document}